\documentclass[usenatbib]{mn2e}
\usepackage{natbib}
\usepackage[dvips]{graphicx}
\usepackage{aas-macros}
\bibpunct[, ]{(}{)}{;}{a}{}{,}
\newcommand{\pct}{\,\%}
\newcommand{\dif}{\mathrm{d}}
\newcommand{\diff}[2]{\frac{\dif#1}{\dif#2}}

\newcommand{\xibd}{\xi_\mathrm{BD}}
\newcommand{\xip}{\xi_\mathrm{pri}}
\newcommand{\xist}{\xi_\mathrm{star}}
\newcommand{\nbod}{N_\mathrm{bod}}
\newcommand{\nsng}{N_\mathrm{sng}}
\newcommand{\nbny}{N_\mathrm{bny}}
\newcommand{\nbd}{N_\mathrm{BD}}
\newcommand{\nst}{N_\mathrm{star}}

\newcommand{\abd}{\alpha_\mathrm{BD}}
\newcommand{\rpop}{\mathcal{R}_\mathrm{pop}}
\newcommand{\rbod}{\mathcal{R}}

\newcommand{\rhbl}{\mathcal{R}_\mathrm{HBL}}

\newcommand{\msun}{M_{\sun}}
\newcommand{\tmsun}{\mbox{$\msun$}}
\newcommand{\mhibd}{m_\mathrm{max,BD}}

\newcommand{\mmin}{m_\mathrm{min}}
\newcommand{\mpri}{m_\mathrm{prim}}

\newcommand{\ftot}{f_\mathrm{tot}}
\newcommand{\fbd}{f_\mathrm{BD}}
\newcommand{\fst}{f_\mathrm{star}}
\newcommand{\oimf}{\mbox{$\mbox{IMF}_\mathrm{obs}$}}
\newcommand{\pimf}{\mbox{$\mbox{IMF}_\mathrm{pri}$}}
\newcommand{\simf}{\mbox{$\mbox{IMF}_\mathrm{sys}$}}
\newcommand{\stdens}{n}
\begin{document}
\title[A discontinuity in the low-mass IMF]{A discontinuity in the low-mass IMF -- the case of high multiplicity}
\author[Thies, Kroupa]{Ingo Thies$^{1}$, Pavel Kroupa$^{1}$\\
$^1$Argelander-Institut f\"ur Astronomie (Sternwarte), Universit\"at Bonn, Auf dem H\"ugel 71, D-53121 Bonn, Germany}
\pagerange{1200 -- 1206}\pubyear{2008}\volume{390}
\maketitle
\begin{abstract}
The empirical binary properties of brown dwarfs (BDs) differ from
those of normal stars suggesting BDs form a separate
population. Recent work by Thies and Kroupa revealed a discontinuity
of the initial mass function (IMF) in the very-low-mass star regime
under the assumption of a low multiplicity of BDs of about 15 per
cent. However, previous observations had suggested that the
multiplicity of BDs may be significantly higher, up to 45 per
cent. This contribution investigates the implication of a high BD
multiplicity on the appearance of the IMF for the Orion Nebula
Cluster, Taurus-Auriga, IC~348 and the Pleiades. We show that the
discontinuity remains pronounced even if the observed MF appears to be
continuous, even for a BD binary fraction as high as 60\pct.
We find no evidence for a variation of the BD IMF with star-forming
conditions. The BD IMF has a power-law index $\abd\approx +0.3$ and
about 2 BDs form per 10 low-mass stars assuming equal-mass pairing of
BDs.
\end{abstract}
\section{Introduction}
\label{sec:introd}
The origin of brown dwarfs (BDs) remains the subject of intense
discussions.  There are two broad ideas on their origin: 1.~the
classical star-like formation scenario of BDs
(e.g. \citealt{AdFa96,PaNo04}), and 2.~BDs and some
very-low-mass stars (VLMSs) form as a separate population (hereafter
named BD-like besides the classical star-like
population) with a different formation history than stars, e.g. as
ejected stellar embryos \citep{ReiCla01,KB03b} or as disrupted wide
binaries \citep{GoWi07,StHuWi07}.  Additionally, the formation of
BDs from Jeans instabilities in high-density filaments near the centre
of a massive star-forming cloud has been recently suggested implying
such BDs to be preferentially located in clusters with strong
gravitational potentials \citep{BCB08}.

The star-like formation scenario fails to reproduce the observed
different binary properties of BDs and stars. Especially the
truncation of the semi-major axis distribution between 10 and 20~AU
for BDs and the different mass-ratio distribution of BDs and stars
\citep{Bouyetal03,Burgetal03,Maetal03,Ketal03,Cloetal03}, as well as
the BD desert \citep{2003IAUS..211..279M,GreLin06},
are difficult to account for if BDs form indistinguishably to
stars. This implies the need of treating BDs as a separate population
to stars.  The assumption of two separate populations would then
require two separate initial mass functions (IMFs) of the individual
bodies of a star cluster.
Although the observed mass function may appear approximately
continuous from the lowest mass BDs up to the highest mass stars
\citep{UKIDSS2007}, a discontinuity in the IMF may be present but be
masked by `hidden' (unresolved) binaries only emerging if the
observed MF is corrected for unresolved multiplicity.  This issue has
been discussed in greater detail in Thies \& Kroupa (2007, hereafter
TK07) for the case of a low multiplicity of 15\pct\ of the
BD-like population.  However, a higher BD multiplicity between 20\pct\
and 45\pct\ had been reported by some authors,
e.g. \citet{JefMax05,BasRei2006}.  This raises the need, dealt with
this contribution, for including higher multiplicities as well as for
an analysis of the general effects of a high multiplicity on the IMF
and on the BD-to-star ratio.

In Section \ref{sec:bdpop} we review the evidence for a separate
BD-like population.  Section \ref{sec:method} briefly introduces the
mathematical method of calculating the IMF including unresolved
binaries. In Section \ref{sec:results} the new results are presented
and compared to those of TK07. The Summary follows in
Section~\ref{sec:summary}.

\section{Brown dwarfs as a separate population}
\label{sec:bdpop}
\subsection{Motivation}
\label{ssec:motiv}

\begin{figure}
\begin{center}
\includegraphics[width=8cm]{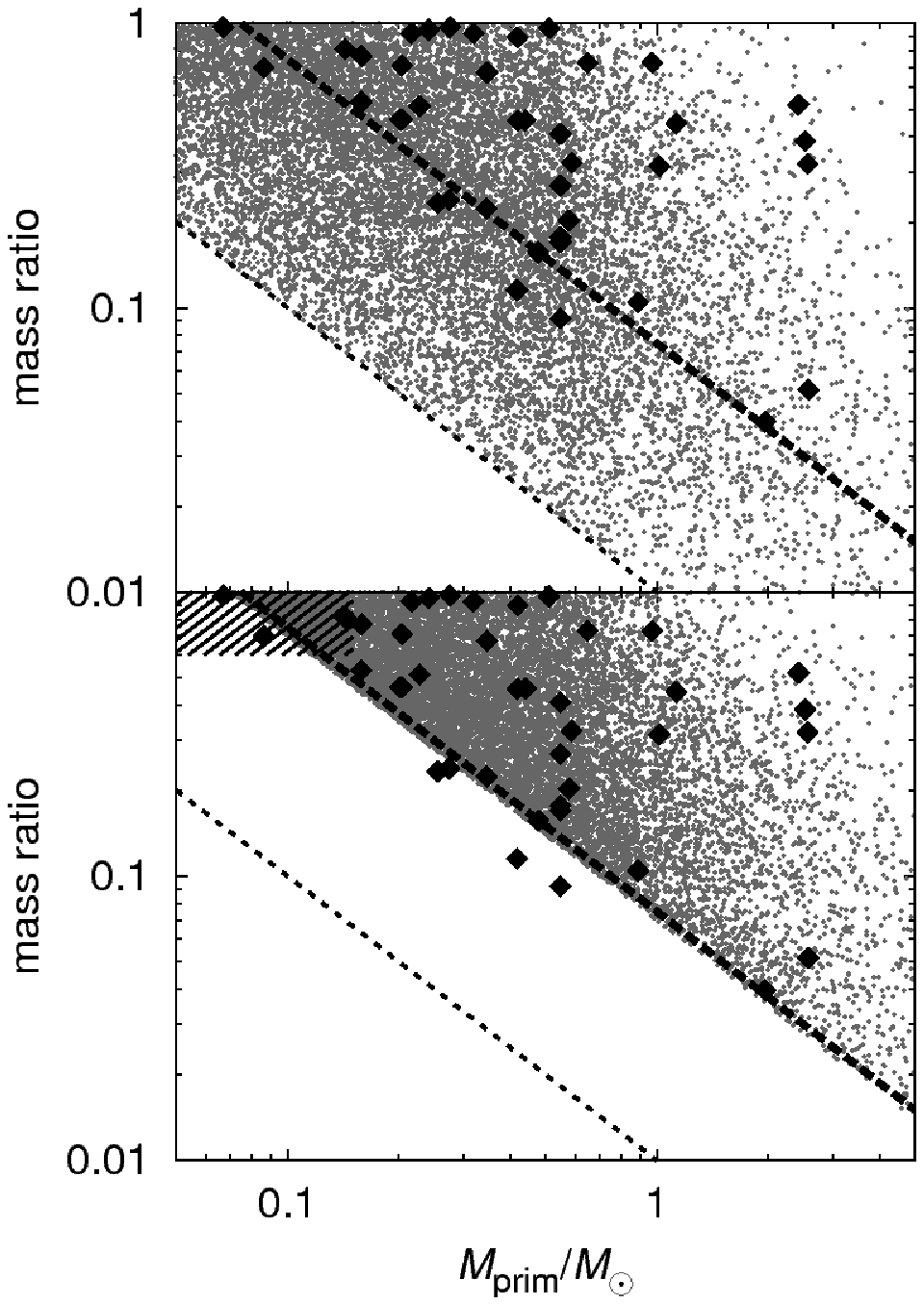}
\caption{\label{q2random}Mass ratio distribution of binaries in a
Monte Carlo sample cluster following the canonical stellar mass
function (\S~\ref{sec:method}). Upper panel: Binaries formed via
random pairing over the complete mass range of a single population of
BDs and stars (dots). The companions have masses down to
0.01~\tmsun\ (thin dashed line) for all kinds of primaries. This is in
contradiction with observations (\citealt{LafJayBra08}, black diamonds).
Lower panel: Binaries from
random pairing within separate populations of BDs between 0.01 and
0.075~\tmsun, and stars above 0.075~\tmsun, i.e. there are no stellar
companions below 0.075~\tmsun\ (thick dashed line).
The resulting mass-ratio distribution fits well to the one observed by
\citet{LafJayBra08} in the stellar regime. The regime of the
(approximately equal-mass companion) BD-like binaries is indicated by
the shaded region.}
\end{center}
\end{figure}

It may be argued that by treating the BDs as a separate population
this forces an IMF discontinuity near the stellar/sub-stellar
mass limit by construction. Indeed, the semi-major axis data
and binary fraction (here used as a simplification for the
multiplicity, neglecting multiples of higher order, \citealt{PPVGoodwinetal})
as a function of the primary mass can be interpreted to
be continuous with no evidence for BDs being a separate population
\citep{BurgetalPPV}.

Given this argumentation, it is essential to describe the
methodology applied in our analysis: We seek one mathematical
formulation which is a unification of the binary population for
G-, K-, M-dwarfs and VLM-stars and BDs. This is found to be
possible for G-, K- and M-dwarfs: thus, for example, G-dwarfs have
mostly K- and M-dwarf companions, and the period-distribution functions
of G-, K- and M-dwarf primaries are indistinguishable. Further, the
mass-ratio and period-distribution functions for G-, K- and M-dwarf
primaries derive from a single birth mass-ratio and period distribution
which does not differentiate according to the mass of the primary
\citep{Ketal03,PPVGoodwinetal}.
One single mathematical model can therewith be written down which treats
G-, K- and M-dwarfs on exactly the same footing -- one can say that G-,
K- and M-dwarf stars mix according to one rule (random pairing from the
IMF at birth).

If BDs are to be introduced into a similar mathematical formulation
which does not differentiate between BDs and stars, then the model
fails, because it leads to (1) a too wide BD period-distribution
function, (2) too many BD binaries, (3) far too many stellar--BD
binaries, and (4) far too few star--star binaries \citep{Ketal03}.
\citet{LafJayBra08} show that the mass-ratio of binaries depends on
the primary-star mass in a way that results in an almost constant
lower mass limit of the companions near 0.075~\tmsun\ for the
Chamaeleon~I star-forming region (see the figures 7 and 12 in their
paper).  This distribution can be reproduced by random pairing over
the stellar mass range, while global random pairing over BDs and stars
yields a different distribution, as shown in Fig. \ref{q2random}.  In
order to avoid these failures, and in particular, in order to
incorporate the BD desert (`stars and BDs don't mix, while G, K and M
dwarfs do mix') into the mathematical formulation of the population,
it is unavoidable to invent special mathematical rules for the
BDs. That is, stars and BDs must be described separately.

TK07 show that  this necessarily implies a discontinuity in
the IMF, given the observational data.
We emphasise that the
observational mass distributions lead to this conclusion, once the
correct mathematical description is incorporated consistently.
However, TK07 assume a rather low binary fraction of the BD-like
population of 15\pct. Some papers \citep{GuWu03,
Kenetal05,Joergens2006a} that report the discovery of close BD
binaries instead conclude that these may imply a significantly higher
binary fraction between 20\pct\ and 45\pct\
\citep{JefMax05,BasRei2006}, the latter being similar to that of stars
in dynamically evolved environments \citep{Kr95b}. It is therefore
useful to re-address the problem TK07 posed by incorporating a larger
BD binary fraction into the analysis.

\subsection{A short review of binarity analysis}
\label{ssec:rev}
This contribution is, like TK07, part of a series on the theoretical
interpretation of observational stellar cluster data.
\citet{KGT91,KTG93,Kr01} showed for the first time that the true
individual-body stellar IMF is changed significantly by correcting for
the bias due to unresolved binary stars.  A detailed study in
\citet{Kr95a,Kr95b,Kr95c}, in \citet*{KPM99}, and in \citet*{KAH01} of
the observed binary properties of field stars, stars in the
Orion Nebula Cluster (ONC), the Pleiades and the Taurus-Auriga
association (TA) led to a thorough understanding of the energy
distribution of binary systems.  This work established that simply
taking observed distributions can lead to wrong interpretations,
unless the counting biases and stellar-dynamical evolution is treated
systematically and consistently; the basis of the argument being that
Newton's laws of motion cannot be ignored.  In a recent paper
\citet{Reietal07} have supported the predictions made by \citet{KPM99}
concerning the binary population in the ONC, by uncovering a radially
dependent binary fraction in nice agreement with the theoretically
expected behaviour.  The late-type stellar binary population is
therewith quite well understood, over the mass range between about
0.2~\tmsun\ and 1.2~\tmsun. The above work has also established the
necessity to correctly dynamically model observed data in order
to arrive at a consistent understanding of the physically relevant
distribution functions.

Brown dwarfs (BDs), which extend the mass scale down to 0.01~\tmsun, have been
added into the theoretical analysis in \citet{KB03a,KB03b,Ketal03}.
This theoretical study of observational data \citep{Cloetal03,Bouyetal03}
showed that BDs cannot be understood to be an extension of the stellar mass
regime (as is often but wrongly stated).
The hypothesis of doing so leads to incompatible
statistics on the star-star, star-BD and BD-BD binary fractions, and on
their energy distributions.
This work showed that BDs must be viewed as a
separate population, and the theoretical suggestion by \citet{ReiCla01},
that BDs are ejected stellar embryos, is one likely explanation
for this. In fact, their proposition logically implies different binary
properties between stars and BDs, because ejected objects cannot have the
same binding energies as not-ejected objects.
Likewise, the model of \citet{GoWi07}, according to which BDs are born in
the outer regions of massive accretion disks, implies them to have
different pairing rules than stars.

TK07 and this contribution are a logical extension of the above findings.
Here we repeat parts of the analysis of TK07 with assumed
BD-like binary fractions up to 60\pct\ as an upper limit. The clusters
we analyse are the ONC (\citealt{Muetal02}),
TA (\citealt{Lu04b,Luetal03a}),
IC~348 \citep{Luetal03b} and the Pleiades based on
data by \citet{Doetal02}, \cite{Moetal03} and the Prosser and Stauffer
Open Cluster Database\footnote{Available at
http://www.cfa.harvard.edu/\symbol{126}stauffer/opencl/}.
The aim is to check whether our previous results are still valid for a
higher binary fraction, and how robust they are for different accounting
of unresolved binary masses.

\section{IMF basics and computational method}\label{sec:method}
The IMFs are constructed from power-law functions similar to that proposed by
\citet{salpeter1955},
\begin{equation}\label{powerimf}
\xi(m)=\diff{n}{m}=k\,m^{-\alpha}\,,
\end{equation}
or in bi-logarithmic form
\begin{equation}\label{powerlog}
\xi_\mathrm{L}(\log_{10}m)=\diff{n}{\log_{10}m}=\ln\left(10\right)\,m\,\xi(m)=k_\mathrm{L}\,m^{1-\alpha}\,,
\end{equation}
where $k$ is a normalisation constant and $k_\mathrm{L}=\ln(10) k$.
While Salpeter found $\alpha\approx2.35$, the canonical stellar
IMF, $\xist$, is constructed as a two-part-power law after \citet{Kr01},
with $\alpha_1=1.3$ for a stellar mass $m<0.5$~\tmsun\ and
$\alpha=2.3$ for higher masses. The substellar IMF, $\xibd$, is taken
to be a single power-law with cluster-dependent exponent $\abd$.

The basic assumption is that a large fraction of binaries remains
unresolved since cluster surveys are often performed with wide-field
surveys with limited resolution.
One may be tempted to use the observed
IMF (hereafter \oimf) as a direct representation of the true IMF of
individual bodies (simply the IMF hereafter). However, the observed IMF
(\oimf) can differ largely from the IMF, especially at
the low-mass end of the population which contains most of the
stellar companions. If the companion has a much
lower mass than its primary, then its light does not contribute much
to the combined luminosity and spectral type, and thus the derived mass
is essentially that of the primary. If, however,
both components have near-equal masses (as expected from random pairing
for very low primary masses), the low-mass region of the \oimf\ may be
depressed even further, since the combined luminosity can be up to
twice the luminosity of the primary alone.
Therefore a fraction of unresolved low-mass binaries is counted as single
stars, maybe even of higher mass, while their companions are omitted,
attenuating the \oimf\ at the lower mass end.

Possible approximations to the \oimf\ are the system IMF (\simf), that
is the IMF as a function of system mass (see equations 6 to 8 in TK07),
and the primary body IMF (\pimf), the IMF as a function of the
primary object mass, $\mpri$,
\begin{equation}\label{defxip}
\xip(\mpri)=\ftot\nbod\int\limits_{\mmin}^{\mpri}\hat{\xi}(\mpri)\hat{\xi}(m)\,\dif{m}\,,
\end{equation}
where $\nbod$ is the total number of objects, $\mmin$ is the minimum
mass of an individual body in the given population,
$\ftot\equiv\nbny/(\nsng+\nbny)$ is the total binary fraction, and
$\hat{\xi}(m)=\xi(m)/\nbod$ is the normalised individual-body IMF.

In TK07 the \simf\ has been used for the fitting process.  However,
one may argue that the mass derived from the system luminosity is
closer to the mass of the primary star since the luminosity is mainly
given by the primary object and the spectral features of the companion
are outshone by those of the primary. Therefore, the \pimf\ has been
used as the workhorse in the current contribution.

To obtain the true IMF from an observed mass distribution a binary
correction has to be applied to each native population (i.e. a
population of objects that share the same formation history) the
cluster consists of. This is done here via the
semi-analytical backward-calculation method and $\chi^2$ minimisation
against the observational data introduced in TK07. It assumes two
native populations with different IMFs, different overall
binary fractions and different mass-ratio distributions (namely the two
extreme cases of random pairing and equal-mass pairing for BDs while
random pairing is always applied to stars).  For each cluster the BD
IMF slopes, the population ratio,
\begin{equation}\label{defrpop}
\rpop=\frac{\nbd}{\nst}\,,
\end{equation}
and the upper
mass limit of the BD-like IMF, $\mhibd$, are to be fitted,
while the lower mass limit of BDs (0.01~\tmsun)
and of the star-like population (0.07~\tmsun) is kept constant. Here,
the number of BD-like and star-like objects is given by
\begin{equation}\label{defnpop}
\begin{array}{rcl}
\nbd&=&\int \xibd(m)\,\dif m\,,\\
\nst&=&\int \xist(m)\,\dif m\,,
\end{array}
\end{equation}
respectively.
The IMFs are then transformed into separate primary
mass functions. Before being compared to the observational MFs
the fitted \pimf\ has been smoothed
by a Gaussian convolution along the mass axis in order to simulate the
error of the mass determination (see TK07 for a more detailed
description). This process is repeated iteratively until $\chi^2$
reaches a minimum.

For the Pleiades the BD data do not constrain the power-law index,
so fixed power-laws with $\abd=0.3$ (the canonical value)
and $\abd=1$ have been used here.
It should be noted that the power-law indices are in rough agreement with
the power-law index $\alpha=0.6$ deduced by \citet{Bouetal98} and
\citet{Moetal03}. Since they use BDs and low-mass stars
up to 0.48~\tmsun\ while only BDs and VLMSs are used in our contribution
these values have to be compared with caution.

The crucial point in performing the binary correction is that the
assumed number and mass range of a native population largely affects
the resulting \pimf. If, for example, only one overall population is
assumed (as in the traditional star-like scenario for BDs and stars)
but there are actually two separate BD-like and star-like populations
with different mass ranges, then the binarity is corrected for wrongly
at the lower mass end of the star-like population since a mixing of
binary components between BDs and stars is assumed that does
not exist in reality. Reversely, the observed (primary) IMF of a cluster
may appear as being continuous while actually consisting of two
populations, because the discontinuity is masked by the interference
of different probability densities of the populations in the transition
or overlap region on the one hand and different binary fractions on the
other, as well as being smeared out by measurement uncertainties.
Thus, an apparently continuous \pimf\ or \simf\ may be related
to a discontinuous IMF or, more precisely, a composite IMF which
can only be revealed by reducing the fraction of unresolved binaries to
insignificance by high-resolution observations.

The magnitude of the discontinuity, measured as the number ratio of
BD-like to star-like objects at the hydrogen-burning mass limit (HBL),
$\rhbl$, is given by
\begin{equation}\label{defrhbl}
\rhbl=\frac{\nbd\left(m\approx0.075\,\msun\right)}{\nst\left(m\approx0.075\,\msun\right)}\,.
\end{equation}
If there is no overlap of the fitted BD-like and the star-like
population (which is actually the case for the ONC and the Pleiades),
$\rhbl$ is calculated from extrapolation of the BD-like IMF to the
HBL. Since $\rhbl$ depends on the binary fraction among each
population, the binary fraction is varied from $\fbd=0$ to $\fbd=0.6$
in order to include even the most extreme BD binary fraction. The
unresolved stellar binary fraction, $\fst$, is set to 0.4 for the ONC,
IC~348 and the Pleiades while that of TA is assumed to be 0.8, as in
TK07.

\section{Results}\label{sec:results}
\subsection{IMF fitting parameters for different BD binary fractions}\label{ssec:imfpars}
\begin{figure}
\begin{center}
\includegraphics[width=8cm]{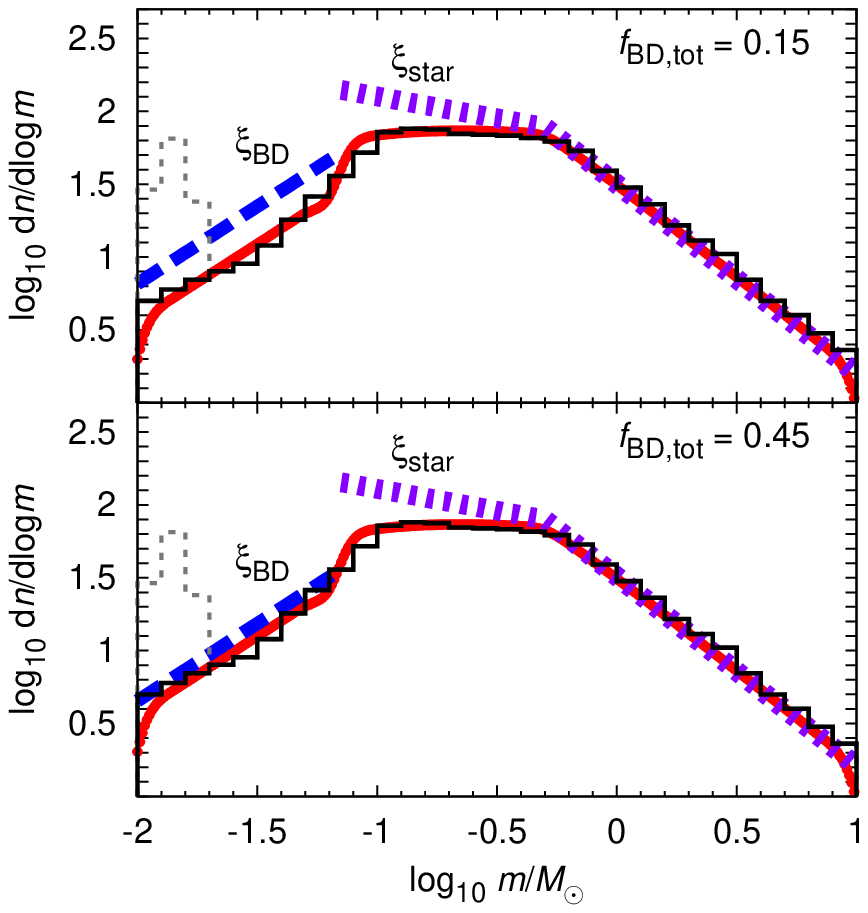}
\caption{\label{imf2tp}The best-fit IMFs for the ONC
(\citealt{Muetal02}, histogram) for $\fbd=0.15$, as displayed in TK07,
and for $\fbd=0.45$, the upper limit deduced by \citet{JefMax05}.  The
primary IMFs (solid curves) are being derived from separated BD-like
(dashed lines) and star-like (dotted lines) populations consistent
with the ejected-embryo hypothesis of \citet{ReiCla01} or the
disk-fragmentation hypothesis of \citet{GoWi07}.  The assumed fraction
of unresolved star-like binaries is $\fst=0.4$.  Despite the high
binary fraction in the lower panel the discontinuity near the
stellar-substellar border is only slightly reduced.  The unequal
binary fractions for different masses and populations mask the
apparent discontinuities of the IMFs in the VLMS region in \pimf\
(solid curves).  The thin dashed histogram refers to a substellar peak
in the data from \citet{Muetal02} which may be due to non-physical
artefacts in the mass-luminosity relation used for the mass
calculation \citep{LaLa03}.}
\end{center}
\end{figure}
\begin{figure}
\begin{center}
\includegraphics[width=8.3cm]{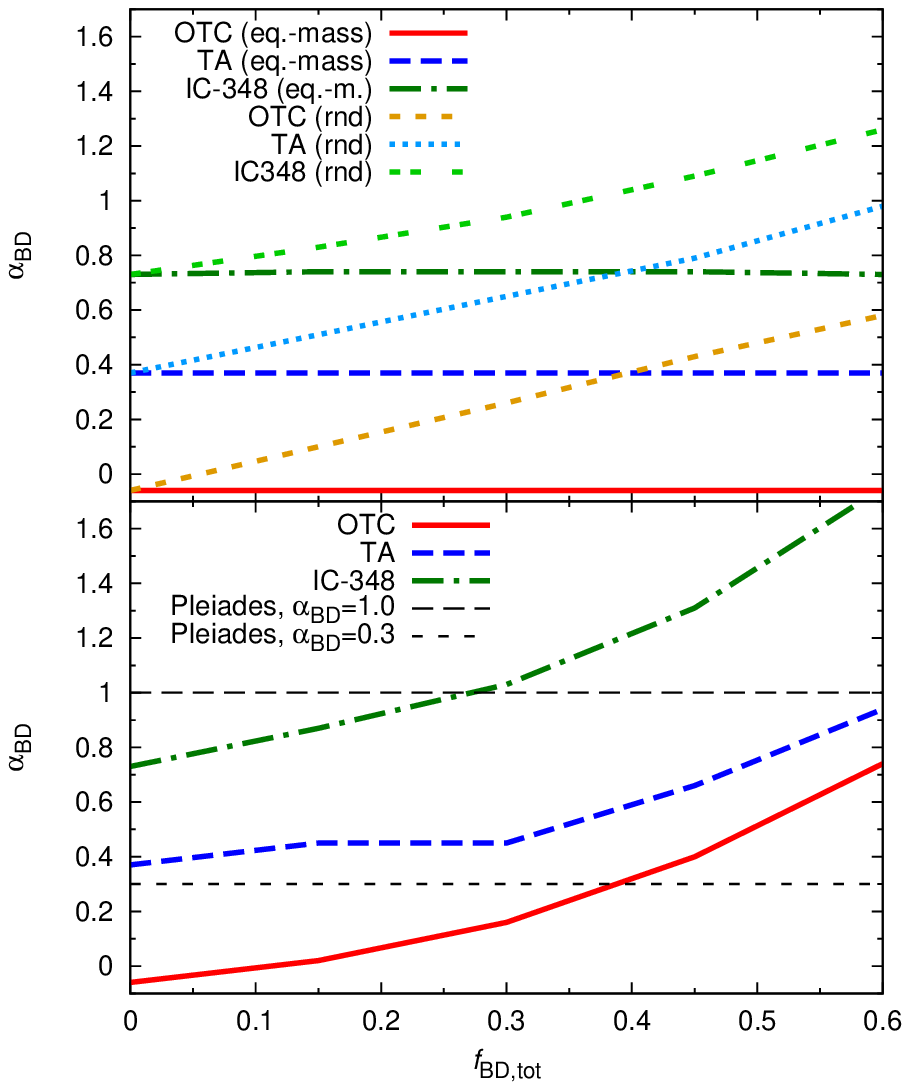}
\caption{\label{fbd_a}{\bf Upper panel}:~The best-fit BD IMF power-law
indices for the ONC (solid line), TA (dashed line), and IC~348
(dash-dotted line) with equal-mass pairing and random pairing of BD
binaries (dotted, narrow-dotted and double-dotted curves,
respectively), as a function of the assumed BD binary fraction,
$\fbd$, fitted via the primary-object mass function for both BDs and
stars (see text).  The upper/lower uncertainty limits of $\abd$ are
about $+0.3/-0.3$ for the ONC, $+0.5/-2.8$ for TA and $+2/-0.6$ for
IC~348. While $\abd$ remains approximately constant for equal-mass
pairing, there is a strong increase with increasing $\fbd$ for the
random pairing case.  {\bf Lower panel:}~The BD IMF power-law indices
for the case of equal-mass pairing, but this time fitted via the BD
system MF (and the stellar primary MF).  The upper/lower limits of
$\abd$ are about $+0.3/-0.3$ for the ONC, $+0.7/-3.5$ for TA and
$+2.5/-0.6$ for IC~348. In contrast to the equal-mass case in the
upper panel, $\abd$ is now increasing with $\fbd$. For comparison, the
fixed $\abd=0.3$ (short-dashed horizontal line) and $\abd=1.0$
(long-dashed horizontal line) for the Pleiades have been added.}
\end{center}
\end{figure}
For illustration, Fig. \ref{imf2tp} shows the fitted BD-like and
star-like IMFs, $\xibd$ and $\xist$ and the
resulting \pimf\ for the ONC for an assumed $\fbd=0.15$ (upper panel)
and $\fbd=0.45$ (lower panel), both using equal-mass pairing for
BD-like binaries and random pairing for star-like ones.
Random pairing means, in this context, pairing two stars selected
by chance from the IMF.
The discontinuity (eq. \ref{defrhbl})
between the BD-like and the
star-like IMF becomes slightly smaller for higher binary fractions while
the BD-like IMF slope remains almost constant. The discontinuity between
both populations is, however, still present.

The top panel of Fig. \ref{fbd_a} shows the dependency of $\abd$ on $\fbd$
for the ONC, TA and IC~348 (the clusters for which $\abd$ has actually been
calculated from $\chi^2$ minimisation) for both equal-mass pairing
and random pairing. The IMFs have been fitted via the \pimf. It should be
noted that $\abd\ge1$ for $\fbd\ga0.4$ for random-pairing in IC~348,
i.e. the turnover of the (bi-logarithmic) IMF in the substellar region
vanishes, although the discontinuity remains (see Section \ref{ssec:rhbl}).
Similarly, the lower panel shows the trends with $\fbd$ in the
case of equal-mass pairing if the BD-like \simf\ is used for fitting.

The most remarkable feature is that $\abd$ remains almost constant
for equal-mass pairing in BD-like binaries. For random-pairing
$\abd$ increases with $\fbd$ in a similar way for all three clusters.
A similar growth is found even for equal-mass pairing if \simf\ is
used for fitting. For comparison, the constant values assumed 
for the Pleiades are shown in the lower panel of Fig.~\ref{fbd_a}
(straight dashed lines at $\abd=0.3$ and $\abd=1.0$).

The fitting of $\mhibd$ yields values slightly below 0.07~\tmsun\ for
the ONC and the Pleiades.  This is probably due to the Gaussian
smearing of $\log m$ that has been used for smoothing the fit.  For TA
and IC~348, however, $\mhibd$ is found to be around 0.1~\tmsun\ and
between 0.15 and 0.23~\tmsun, respectively. Furthermore, our
results for the best-fit $\rpop$, the population ratio, and the
magnitude of the discontinuity can be summarised as follows: For
$\fbd=0$ the best-fit $\rpop$ is about 0.07 for the ONC and the
Pleiades while it is about~0.15 for TA and IC~348. It increases for
larger $\fbd$, reaching about twice these values for the extreme
binarity of $\fbd=0.6$. That is, if a realistic value of
$\fbd\approx0.2$ is assumed, we expect about~1 BD-like body per~10
star-like ones for the ONC and the Pleiades, and about~1 BD-like body
per~5 star-like bodies for the others. This result is remarkable given
that e.g. \citet{SHC04} state a higher BD-to-star ratio for the
ONC than for TA and IC~348. The result can be interpreted as a
consequence of the large mass overlap of the BD-like and the star-like
regime in IC~348 (and a moderate overlap in TA), i.e. that many
BD-like bodies are in actually very-low mass stars and thus are
counted as normal stars.  Another issue is whether the substellar peak
in the ONC MF (see the thin dashed histogram in Fig. \ref{imf2tp}) is
an artefact \citep{LaLa03} or a real feature. In the latter case,
$\rpop$ would be significantly higher (about 75\pct, given the
histogram data) for the ONC than suggested by our results.

One may criticise the way of assigning a mass to an observed system.
In TK07 the model-observed IMF has been created by simply
adding the masses of all components, i.e. $\oimf=\simf$. Because the
observed data are being derived from
luminosity functions rather than from mass functions, the correct way would
be to convert luminosities
into masses via the mass-luminosity relation (MLR). This would require
rather complicated calculations because full-scale modelling would
involve age-spreads and age-dependent MLRs with very significant
uncertainties \citep{WuTscha03}.

Instead, a simpler way to at least embrace the real relation is to
repeat the analysis (or parts of it) by using the primary mass instead
of the system mass. This corresponds to the extreme case that the
contribution of less-massive companions is negligible.  This method
has been used in the present contribution with similar results
as in TK07. In addition, similar calculations have been made
for a substellar system IMF and for a stellar primary IMF, for
equal-mass pairing of BDs, and for random pairing of stars.

\subsection{The discontinuity in the low-mass IMF}
\label{ssec:rhbl}

\begin{figure}
\begin{center}
\includegraphics[width=8.3cm]{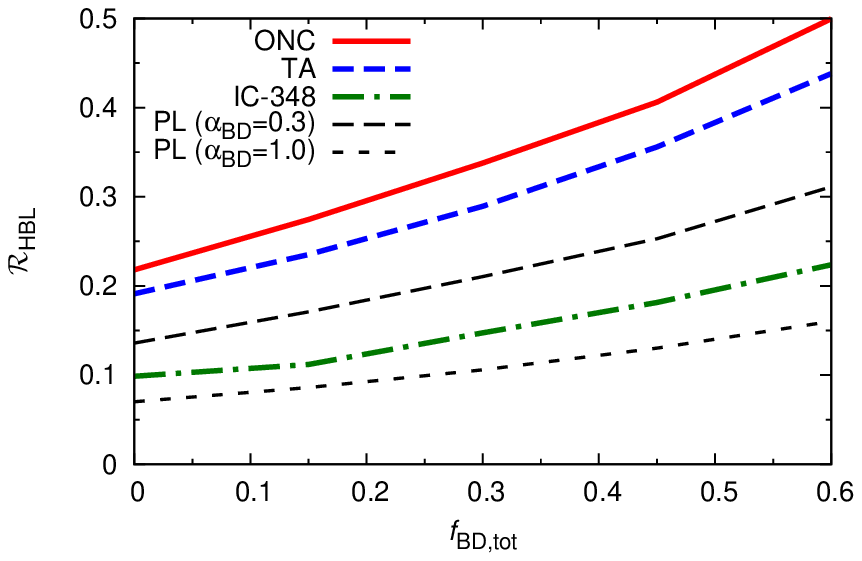}
\caption{\label{rhbl}The ratio of BD-like to star-like bodies at the
hydrogen-burning mass limit (HBL) as a function of the BD-like
binarity, $\fbd$, for the ONC (solid curve), TA (dashed curve), IC~348
(dash-dotted curve), as well as for the Pleiades for $\abd=0.3$ (thin
dashed curve) and $\abd=1.0$ (thin dotted curve). For all clusters
there is a similar trend towards a higher $\rhbl$ with increasing
binary fraction. But even for the highest plausible binary fraction
$\rhbl<0.5$. A continuous IMF would require $\rhbl\equiv1$.  }
\end{center}
\end{figure}

A measure for the discontinuity at the HBL, $\rhbl$, is given
by eq. \ref{defrhbl}. For a continuous IMF $\rhbl\equiv1$, while
values significantly different from~1 indicate a discontinuity.
Fig.~\ref{rhbl} displays $\rhbl$ as a function of $\fbd$. For all
clusters $\rhbl$ shows a similar steady increase. The uncertainties
(not shown in the graph) can be estimated from those of $\abd$ and are
about $\pm{40}\pct$ for each value.  Thus the discontinuity between
the BD-like and the star-like IMF persists for both the system and the
primary-body IMF for the BD-like population. It is largest
(i.e. $\rhbl$ is smallest) for $\fbd=0$.  The results from TK07 show
that this even holds true if the system IMF fit is applied to the
stellar population.

\subsection{IMF slope and BD-to-star ratio in relation to the stellar density}
\label{ssec:Rmass}

\begin{figure}
\begin{center}
\includegraphics[width=8cm]{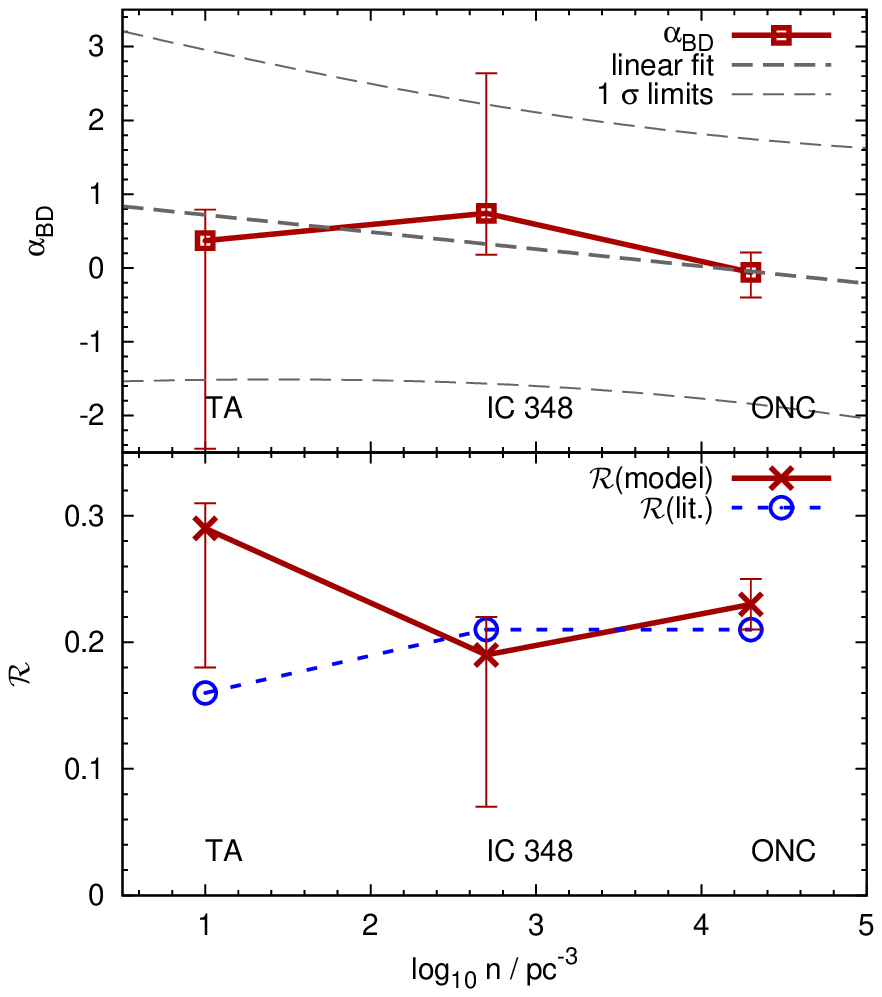}
\caption{\label{abd}The power-law index $\abd$ and the BD-to-star
ratio $\rbod$ from our model (assuming equal-mass pairing for BDs) and
from the literature are plotted in dependence of the logarithmic
central stellar density of TA, IC~348 and the ONC. The references for
$\rbod$ are \citet{Lu06} for TA, \citet{PSZ03} for IC~348, and
\citet{SHC04} for the ONC.}
\end{center}
\end{figure}

Certain theories of BD formation (e.g.  \citealt{BCB08})
suggest a dependency of the rate of BD formation on the star-cluster
density.  Correlating the BD IMF index, $\abd$, and the BD-to-star
ratio, $\rbod$, against the stellar density, $\stdens$, may uncover
such expected dependencies.  For TA, $\stdens=1-10$ stars per
$\mathrm{pc}^{3}$ \citep{Martinetal01},
$\stdens\approx500\,\mathrm{pc}^{-3}$ for IC~348
\citep{Ducheneetal99}, and $\stdens\approx20000\,\mathrm{pc}^{-3}$ for
the ONC \citep{HillHart98}.  The values of $\abd$ found in this study
are plotted against $\log_{10}\stdens$ in Fig. \ref{abd} (upper
panel). In addition, the BD-to-star ratio, $\rbod$, is shown in the
lower panel, where
\begin{equation}\label{defrbod}
\rbod=\frac{N(0.02\,\msun \le m  \le0.075\,\msun)}{N(0.15\,\msun\le m \le 1\msun)}\,.
\end{equation}
The mass limits are chosen in accordance with \citet{Ketal03} and
\citet{TK07}.  The crosses connected with solid lines show the results
of our modelling while the open circles with dashed lines are values
taken from \citet{Lu06} (TA), \citet{PSZ03} (IC~348), and
\citet{SHC04} (ONC).

For $\abd$ a regression line has been calculated. However, only three
clusters have been analysed in this study, and there are large
uncertainties. Especially for TA and IC~348 the confidence range is rather
large here. Thus, the linear fit is only poorly constrained and well in
agreement with a constant $\abd$.
Furthermore, $\rbod$ also does not show a significant trend with
increasing stellar density. From our analysis (Fig. \ref{abd}) it follows
that $\abd\approx0.3$ and $\rbod\approx0.2$ for equal-mass pairing of BDs.

\section{Summary}
\label{sec:summary}
A discontinuity in the IMF near the hydrogen burning mass limit
appears if the binary properties of BDs and VLMSs on the one hand, and
of stars on the other, are taken into account carefully when inferring
the true underlying single-object IMF. This implies that BDs and some
VLMSs need to be viewed as arising from a somewhat different formation
channel than the stellar formation channel, but this result has been
obtained by TK07 under the assumption that BDs have a binary fraction
of only~15\pct. A higher binary fraction may close the gap between the
stellar and the BD IMF. We refer to BDs and those VLMSs formed
according to the putative BD channel as ``BD-like'' bodies, whereas
stars and those BDs formed according to the stellar channel as
star-like. The BD-like channel remains unknown in detail, but
theoretical ideas have emerged (Sections~\ref{sec:introd},
\ref{ssec:rev}, and~\ref{ssec:Rmass}).

Here we have extended the analysis of TK07 for BD-like binary
fractions up to 60\pct\ for the Orion Nebula Cluster, the
Taurus-Auriga association, IC~348 and the Pleiades by using slight
modifications of the techniques introduced in TK07.

As a main result, we found that the discontinuity that comes about by
treating BDs/VLMSs and stars consistently in terms of their observed
multiplicity properties remains even for the highest BD binary fraction.
These results suggest that the BD binary fraction, $\fbd$, is not the
dominant origin of the discontinuity in the IMF, and that, consequently,
two separate IMFs need to be introduced.

It is re-emphasised that by seeking to mathematically
describe the BD and stellar population in terms of the relevant mass-
and binary distribution functions, it is unavoidable to mathematically
separate BDs and VLMSs from stars. The two resulting mass
distributions do not join at the transition mass near 0.08~\tmsun. The
physical interpretation of this logically stringent result is that BDs
and VLMSs follow a different formation history or channel than
stars. This result is obtained independently by theoretical
consideration of star-formation processes \citep{ReiCla01,GoWi07,
StHuWi07,BCB08}.

With this contribution we have quantified how the power-law index of
the BD-like IMF and the BD-to-star ratio changes with varying binary
fraction of BD-like bodies. The BD-like power-law index,
$\abd\approx0.3$, remains almost constant if equal-mass pairing of
BD-like binaries is assumed, while $\abd$ increases somewhat with
increasing $\fbd$ in the case of random pairing over the BD-like mass
range. All values of $\abd$ are between $-0.1$ and~$+1.3$.
We also find that although the stellar density differs from a few
stars per $\mathrm{pc}^3$ (TA) to about 20000 stars per
$\mathrm{pc}^3$, the resulting $\abd$ is constant within the
uncertainties. Similarly, the BD-to-star ratio does not show a trend
with increasing stellar density. This suggests the
star-formation and BD-formation outcome to be rather universal at
least within the range of densities probed here.

\section*{Acknowledgements}
This project is supported by DFG grant KR1635/12-1.


\begin{thebibliography}{}

\bibitem[\protect\citeauthoryear{{Adams} \& {Fatuzzo}}{{Adams} \&
  {Fatuzzo}}{1996}]{AdFa96}
{Adams} F.~C.,  {Fatuzzo} M.,  1996, \apj, 464, 256

\bibitem[\protect\citeauthoryear{{Basri} \& {Reiners}}{{Basri} \&
  {Reiners}}{2006}]{BasRei2006}
{Basri} G.,  {Reiners} A.,  2006, \aj, 132, 663

\bibitem[\protect\citeauthoryear{{Bonnell}, {Clark} \& {Bate}}{{Bonnell}
  et~al.}{2008}]{BCB08}
{Bonnell} I.~A.,  {Clark} P.~C.,    {Bate} M.~R.,  2008, \mnras, 389, 1556

\bibitem[\protect\citeauthoryear{{Bouvier}, {Stauffer}, {Martin}, {Barrado y
  Navascues}, {Wallace} \& {Bejar}}{{Bouvier} et~al.}{1998}]{Bouetal98}
{Bouvier} J.,  {Stauffer} J.~R.,  {Martin} E.~L.,  {Barrado y Navascues} D.,
  {Wallace} B.,    {Bejar} V.~J.~S.,  1998, \aap, 336, 490

\bibitem[\protect\citeauthoryear{{Bouy}, {Brandner}, {Mart{\'{\i}}n},
  {Delfosse}, {Allard} \& {Basri}}{{Bouy} et~al.}{2003}]{Bouyetal03}
{Bouy} H.,  {Brandner} W.,  {Mart{\'{\i}}n} E.~L.,  {Delfosse} X.,  {Allard}
  F.,    {Basri} G.,  2003, \aj, 126, 1526

\bibitem[\protect\citeauthoryear{{Burgasser}, {Kirkpatrick}, {Reid}, {Brown},
  {Miskey} \& {Gizis}}{{Burgasser} et~al.}{2003}]{Burgetal03}
{Burgasser} A.~J.,  {Kirkpatrick} J.~D.,  {Reid} I.~N.,  {Brown} M.~E.,
  {Miskey} C.~L.,    {Gizis} J.~E.,  2003, \apj, 586, 512

\bibitem[\protect\citeauthoryear{{Burgasser}, {Reid}, {Siegler}, {Close},
  {Allen}, {Lowrance} \& {Gizis}}{{Burgasser} et~al.}{2007}]{BurgetalPPV}
{Burgasser} A.~J.,  {Reid} I.~N.,  {Siegler} N.,  {Close} L.,  {Allen} P.,
  {Lowrance} P.,    {Gizis} J.,  2007, in {Reipurth} B.,  {Jewitt} D.,   {Keil}
  K.,  eds, Protostars and Planets V
{Univ. Arizona Press, Tucson}, p. 427

\bibitem[\protect\citeauthoryear{{Close}, {Siegler}, {Freed} \&
  {Biller}}{{Close} et~al.}{2003}]{Cloetal03}
{Close} L.~M.,  {Siegler} N.,  {Freed} M.,    {Biller} B.,  2003, \apj, 587,
  407

\bibitem[\protect\citeauthoryear{{Dobbie}, {Pinfield}, {Jameson} \&
  {Hodgkin}}{{Dobbie} et~al.}{2002}]{Doetal02}
{Dobbie} P.~D.,  {Pinfield} D.~J.,  {Jameson} R.~F.,    {Hodgkin} S.~T.,  2002,
  \mnras, 335, L79

\bibitem[\protect\citeauthoryear{{Duch{\^e}ne}, {Bouvier} \&
  {Simon}}{{Duch{\^e}ne} et~al.}{1999}]{Ducheneetal99}
{Duch{\^e}ne} G.,  {Bouvier} J.,    {Simon} T.,  1999, \aap, 343, 831

\bibitem[\protect\citeauthoryear{{Goodwin}, {Kroupa}, {Goodman} \&
  {Burkert}}{{Goodwin} et~al.}{2007}]{PPVGoodwinetal}
{Goodwin} S.~P.,  {Kroupa} P.,  {Goodman} A.,    {Burkert} A.,  2007, in
  {Reipurth} B.,  {Jewitt} D.,   {Keil} K.,  eds, Protostars and Planets V
{Univ. Arizona Press, Tucson}, pp 133

\bibitem[\protect\citeauthoryear{{Goodwin} \& {Whitworth}}{{Goodwin} \&
  {Whitworth}}{2007}]{GoWi07}
{Goodwin} S.~P.,  {Whitworth} A.,  2007, \aap, 466, 943

\bibitem[\protect\citeauthoryear{{Grether} \& {Lineweaver}}{{Grether} \&
  {Lineweaver}}{2006}]{GreLin06}
{Grether} D.,  {Lineweaver} C.~H.,  2006, \apj, 640, 1051

\bibitem[\protect\citeauthoryear{{Guenther} \& {Wuchterl}}{{Guenther} \&
  {Wuchterl}}{2003}]{GuWu03}
{Guenther} E.~W.,  {Wuchterl} G.,  2003, \aap, 401, 677

\bibitem[\protect\citeauthoryear{{Hillenbrand} \& {Hartmann}}{{Hillenbrand} \&
  {Hartmann}}{1998}]{HillHart98}
{Hillenbrand} L.~A.,  {Hartmann} L.~W.,  1998, \apj, 492, 540

\bibitem[\protect\citeauthoryear{{Jeffries} \& {Maxted}}{{Jeffries} \&
  {Maxted}}{2005}]{JefMax05}
{Jeffries} R.~D.,  {Maxted} P.~F.~L.,  2005, Astronomische Nachrichten, 326,
  944

\bibitem[\protect\citeauthoryear{{Joergens}}{{Joergens}}{2006}]{Joergens2006a}
{Joergens} V.,  2006, \aap, 446, 1165

\bibitem[\protect\citeauthoryear{{Kenyon}, {Jeffries}, {Naylor}, {Oliveira} \&
  {Maxted}}{{Kenyon} et~al.}{2005}]{Kenetal05}
{Kenyon} M.~J.,  {Jeffries} R.~D.,  {Naylor} T.,  {Oliveira} J.~M.,    {Maxted}
  P.~F.~L.,  2005, \mnras, 356, 89

\bibitem[\protect\citeauthoryear{{Kroupa}}{{Kroupa}}{1995a}]{Kr95a}
{Kroupa} P.,  1995a, \mnras, 277, 1491

\bibitem[\protect\citeauthoryear{{Kroupa}}{{Kroupa}}{1995b}]{Kr95c}
{Kroupa} P.,  1995b, \mnras, 277, 1522

\bibitem[\protect\citeauthoryear{{Kroupa}}{{Kroupa}}{1995c}]{Kr95b}
{Kroupa} P.,  1995c, \mnras, 277, 1507

\bibitem[\protect\citeauthoryear{{Kroupa}}{{Kroupa}}{2001}]{Kr01}
{Kroupa} P.,  2001, \mnras, 322, 231

\bibitem[\protect\citeauthoryear{{Kroupa}, {Aarseth} \& {Hurley}}{{Kroupa}
  et~al.}{2001}]{KAH01}
{Kroupa} P.,  {Aarseth} S.,    {Hurley} J.,  2001, \mnras, 321, 699

\bibitem[\protect\citeauthoryear{{Kroupa} \& {Bouvier}}{{Kroupa} \&
  {Bouvier}}{2003a}]{KB03b}
{Kroupa} P.,  {Bouvier} J.,  2003a, \mnras, 346, 369

\bibitem[\protect\citeauthoryear{{Kroupa} \& {Bouvier}}{{Kroupa} \&
  {Bouvier}}{2003b}]{KB03a}
{Kroupa} P.,  {Bouvier} J.,  2003b, \mnras, 346, 343

\bibitem[\protect\citeauthoryear{{Kroupa}, {Bouvier}, {Duch{\^ e}ne} \&
  {Moraux}}{{Kroupa} et~al.}{2003}]{Ketal03}
{Kroupa} P.,  {Bouvier} J.,  {Duch{\^ e}ne} G.,    {Moraux} E.,  2003, \mnras,
  346, 354

\bibitem[\protect\citeauthoryear{{Kroupa}, {Gilmore} \& {Tout}}{{Kroupa}
  et~al.}{1991}]{KGT91}
{Kroupa} P.,  {Gilmore} G.,    {Tout} C.~A.,  1991, \mnras, 251, 293

\bibitem[\protect\citeauthoryear{{Kroupa}, {Petr} \& {McCaughrean}}{{Kroupa}
  et~al.}{1999}]{KPM99}
{Kroupa} P.,  {Petr} M.~G.,    {McCaughrean} M.~J.,  1999, New Astronomy, 4,
  495

\bibitem[\protect\citeauthoryear{{Kroupa}, {Tout} \& {Gilmore}}{{Kroupa}
  et~al.}{1993}]{KTG93}
{Kroupa} P.,  {Tout} C.~A.,    {Gilmore} G.,  1993, \mnras, 262, 545

\bibitem[\protect\citeauthoryear{{Lada} \& {Lada}}{{Lada} \&
  {Lada}}{2003}]{LaLa03}
{Lada} C.~J.,  {Lada} E.~A.,  2003, \araa, 41, 57

\bibitem[\protect\citeauthoryear{{Lafreni{\`e}re}, {Jayawardhana}, {Brandeker},
  {Ahmic} \& {van Kerkwijk}}{{Lafreni{\`e}re} et~al.}{2008}]{LafJayBra08}
{Lafreni{\`e}re} D.,  {Jayawardhana} R.,  {Brandeker} A.,  {Ahmic} M.,    {van
  Kerkwijk} M.~H.,  2008, \apj, 683, 844

\bibitem[\protect\citeauthoryear{{Lodieu}, {Pinfield}, {Leggett}, {Jameson},
  {Mortlock}, {Warren}, {Burningham}, {Lucas} \& {et al.}}{{Lodieu}
  et~al.}{2007}]{UKIDSS2007}
{Lodieu} N.,  {Pinfield} D.~J.,  {Leggett} S.~K.,  {Jameson} R.~F.,  {Mortlock}
  D.~J.,  {Warren} S.~J.,  {Burningham} B.,  {Lucas} P.~W.,    {et al.} 2007,
  \mnras, 379, 1423

\bibitem[\protect\citeauthoryear{{Luhman}}{{Luhman}}{2004}]{Lu04b}
{Luhman} K.~L.,  2004, \apj, 617, 1216

\bibitem[\protect\citeauthoryear{{Luhman}}{{Luhman}}{2006}]{Lu06}
{Luhman} K.~L.,  2006, \apj, 645, 676

\bibitem[\protect\citeauthoryear{{Luhman}, {Brice{\~n}o}, {Stauffer},
  {Hartmann}, {Barrado y Navascu{\'e}s} \& {Caldwell}}{{Luhman}
  et~al.}{2003}]{Luetal03a}
{Luhman} K.~L.,  {Brice{\~n}o} C.,  {Stauffer} J.~R.,  {Hartmann} L.,  {Barrado
  y Navascu{\'e}s} D.,    {Caldwell} N.,  2003, \apj, 590, 348

\bibitem[\protect\citeauthoryear{{Luhman}, {Stauffer}, {Muench}, {Rieke},
  {Lada}, {Bouvier} \& {Lada}}{{Luhman} et~al.}{2003}]{Luetal03b}
{Luhman} K.~L.,  {Stauffer} J.~R.,  {Muench} A.~A.,  {Rieke} G.~H.,  {Lada}
  E.~A.,  {Bouvier} J.,    {Lada} C.~J.,  2003, \apj, 593, 1093

\bibitem[\protect\citeauthoryear{{Mart{\'{\i}}n}, {Barrado y Navascu{\'e}s},
  {Baraffe}, {Bouy} \& {Dahm}}{{Mart{\'{\i}}n} et~al.}{2003}]{Maetal03}
{Mart{\'{\i}}n} E.~L.,  {Barrado y Navascu{\'e}s} D.,  {Baraffe} I.,  {Bouy}
  H.,    {Dahm} S.,  2003, \apj, 594, 525

\bibitem[\protect\citeauthoryear{{Mart{\'{\i}}n}, {Dougados}, {Magnier},
  {M{\'e}nard}, {Magazz{\`u}}, {Cuillandre} \& {Delfosse}}{{Mart{\'{\i}}n}
  et~al.}{2001}]{Martinetal01}
{Mart{\'{\i}}n} E.~L.,  {Dougados} C.,  {Magnier} E.,  {M{\'e}nard} F.,
  {Magazz{\`u}} A.,  {Cuillandre} J.-C.,    {Delfosse} X.,  2001, \apjl, 561,
  L195

\bibitem[\protect\citeauthoryear{{McCarthy}, {Zuckerman} \&
  {Becklin}}{{McCarthy} et~al.}{2003}]{2003IAUS..211..279M}
{McCarthy} C.,  {Zuckerman} B.,    {Becklin} E.~E.,  2003, in {Mart{\'{\i}}n}
  E.,  ed., Brown Dwarfs, IAU Symposium 211,
ASP: San Francisco, pp 279--+

\bibitem[\protect\citeauthoryear{{Moraux}, {Bouvier}, {Stauffer} \&
  {Cuillandre}}{{Moraux} et~al.}{2003}]{Moetal03}
{Moraux} E.,  {Bouvier} J.,  {Stauffer} J.~R.,    {Cuillandre} J.-C.,  2003,
  \aap, 400, 891

\bibitem[\protect\citeauthoryear{{Muench}, {Lada}, {Lada} \& {Alves}}{{Muench}
  et~al.}{2002}]{Muetal02}
{Muench} A.~A.,  {Lada} E.~A.,  {Lada} C.~J.,    {Alves} J.,  2002, \apj, 573,
  366

\bibitem[\protect\citeauthoryear{{Padoan} \& {Nordlund}}{{Padoan} \&
  {Nordlund}}{2004}]{PaNo04}
{Padoan} P.,  {Nordlund} {\AA}.,  2004, \apj, 617, 559

\bibitem[\protect\citeauthoryear{{Preibisch}, {Stanke} \&
  {Zinnecker}}{{Preibisch} et~al.}{2003}]{PSZ03}
{Preibisch} T.,  {Stanke} T.,    {Zinnecker} H.,  2003, \aap, 409, 147

\bibitem[\protect\citeauthoryear{{Reipurth} \& {Clarke}}{{Reipurth} \&
  {Clarke}}{2001}]{ReiCla01}
{Reipurth} B.,  {Clarke} C.,  2001, \aj, 122, 432

\bibitem[\protect\citeauthoryear{{Reipurth}, {Guimar{\~a}es}, {Connelley} \&
  {Bally}}{{Reipurth} et~al.}{2007}]{Reietal07}
{Reipurth} B.,  {Guimar{\~a}es} M.~M.,  {Connelley} M.~S.,    {Bally} J.,
  2007, \aj, 134, 2272

\bibitem[\protect\citeauthoryear{{Salpeter}}{{Salpeter}}{1955}]{salpeter1955}
{Salpeter} E.~E.,  1955, \apj, 121, 161

\bibitem[\protect\citeauthoryear{{Slesnick}, {Hillenbrand} \&
  {Carpenter}}{{Slesnick} et~al.}{2004}]{SHC04}
{Slesnick} C.~L.,  {Hillenbrand} L.~A.,    {Carpenter} J.~M.,  2004, \apj, 610,
  1045

\bibitem[\protect\citeauthoryear{{Stamatellos}, {Hubber} \&
  {Whitworth}}{{Stamatellos} et~al.}{2007}]{StHuWi07}
{Stamatellos} D.,  {Hubber} D.~A.,    {Whitworth} A.~P.,  2007, \mnras, 382, L30

\bibitem[\protect\citeauthoryear{{Thies} \& {Kroupa}}{{Thies} \&
  {Kroupa}}{2007}]{TK07}
{Thies} I.,  {Kroupa} P.,  2007, \apj, 671, 767

\bibitem[\protect\citeauthoryear{{Wuchterl} \& {Tscharnuter}}{{Wuchterl} \&
  {Tscharnuter}}{2003}]{WuTscha03}
{Wuchterl} G.,  {Tscharnuter} W.~M.,  2003, \aap, 398, 1081

\end{thebibliography}
\end{document}